# Deep Learning for Design and Retrieval of Nano-photonic Structures

Itzik Malkiel[1]*, Achiya Nagler [2]*, Uri Arieli[2], Michael Mrejen[2]*, Uri Arieli[2] Lior Wolf [1] and Haim Suchowski[2§]

[1]School of Computer Science, Faculty of Exact Sciences, Tel Aviv University, Tel Aviv 69978, Israel

[2]School of Physics and Astronomy, Faculty of Exact Sciences, Tel Aviv University, Tel Aviv 69978, Israel

§Correspondent author: haimsu@post.tau.ac.il

*These authors contributed equally to this work

**Our visual perception of our surroundings is ultimately limited by the diffraction-limit, which stipulates that optical information smaller than roughly half the illumination wavelength is not retrievable. Over the past decades, many breakthroughs have led to unprecedented imaging capabilities beyond the diffraction-limit, with applications in biology and nanotechnology. In this context, nano-photonics has revolutionized the field of optics in recent years by enabling the manipulation of light-matter interaction with subwavelength structures (1-3). However, despite the many advances in this field, its impact and penetration in our daily life has been hindered by a convoluted and iterative process, cycling through modeling, nanofabrication and nano-characterization. The fundamental reason is the fact that not only the prediction of the optical response is very time consuming and requires solving Maxwell's equations with dedicated numerical packages (4-6). But, more significantly, the inverse problem, i.e. designing a nanostructure with an on-demand optical response, is currently a prohibitive task even with the most advanced numerical tools due to the high non-linearity of the problem (7-8). Here, we harness the power of Deep Learning, a new path in modern machine learning, and show its ability to predict the geometry of nanostructures based solely on their far-field response. This approach also addresses in a direct way the currently inaccessible inverse problem breaking the ground for on-demand design of optical response with applications such as sensing, imaging and also for Plasmon's mediated cancer thermotherapy.**

While computer science has been harnessed to address the diffraction limit in imaging and characterization on one hand (super-resolution techniques such as PALM and STORM techniques and more (9-12)) and to assist with the design process on the other hand (13-19) to date no computational technique is capable of addressing both aspects in an integrated manner. Here, we present an integrated deep learning (DL) approach and show how deep neural networks

(DNNs) can streamline the design process and provide a unique, robust, time-efficient and accurate characterization capability of complex nanostructures based on their far-field optical responses. The complexity of the DNN is able to deal with the high nonlinearity of the inference task and creates a model that holds a bi-directional knowledge. We show that the DL approach is not only able to predict the spectral response of nanostructures with high accuracy, but is also able to address the *inverse* problem and provide a single nanostructure's design, geometry and dimension, for a targeted optical response for both polarizations. This approach offers a path for rapid design of optimal nanostructures for targeted chemicals and bio-molecules which are critical for sensing and integrated spectroscopy.

Big Data and modern machine learning have revolutionized computer science in the past few years. Among the most promising and successful machine learning techniques, DL has emerged as a very powerful method that has achieved state-of-the-art results in various tasks, including computer vision (20), speech recognition (21), natural language processing (22), face recognition and others (23). Inspired by the layered and hierarchical human brain's deep architecture, DL uses multiple layers of non-linear transformation to model high-level abstraction in data. DL has also been successfully employed in research areas beyond computer science, such as in particle Physics (24), Ultra cold science (25), Condensed matter (26), chemical Physics (27) and conventional microscopy (28, 29). While it is common practice in deep learning to separate different problems and to train multiple separate networks for each problem, here we introduce a deep learning architecture that is applied to the design and characterization of metal-dielectric sub-wavelength nano-particles and show that the approach of training a bidirectional network that goes from the optical response spectrum to the nanoparticle geometry and back is significantly more effective for both the design and characterization tasks. This approach is breaking new ground for direct on-demand engineering of plasmonics structures for application in sensing, targeted therapy and more. Moreover, the predictive capability of the DL model also holds great promise for multivariate characterization of nanostructures beyond the diffraction limit.

To demonstrate the paradigm shift enabled by our deep learning approach, we consider the interaction of light with sub-wavelength structures such as plasmonic nanostructures and metamaterials and composite layered metallic nanostructures embedded in dielectric, allowing the control of the properties of the outgoing light (30). Figure 1A illustrates such optical response for both polarizations of the electromagnetic field. The interaction of white light (which contains all the colors) with an array of metallic subwavelength geometries fabricated on a glass substrate, causes partial color transmission due to absorption, which can be different for each polarization of the electromagnetic field. However, due to the diffraction limit, these subwavelength nanostructures cannot be observed by a conventional microscope.

Predicting the far field optical response for a defined nanostructure geometry and composition involves solving the full set of Maxwell equations at each location in space and for each wavelength. However, whereas the far-field spectrum is directly connected to the nanostructure geometry, the 'inverse' problem of inferring the nanoscale geometry from a measured or desired far-field spectrum cannot be obtained (See Fig. 1B). For a simple nanostructure, exhibiting single resonance peaks in each polarization, one can solve it semi-analytically or in an intuitive manner (31), however for a general spectral response associated with more complex geometries, no solution is known. Shallow neural networks, evolutionary algorithms and linear regression (13, 15, 32) have gained some success in solving such a task. However, current techniques are still limited in accuracy and practical feasibility and fall short in modeling of nonlinear problems with high complexity of the underlying physical processes (Fig. 1C). To date, none of these approaches can deal efficiently with the inverse problem and it still takes many cycles of trial and error of modeling and characterization to predict or design a nanostructure, for a desired or measured far-field optical spectral response. It is crucial to stress that the learning phase in DNN is a non recurring effort, meaning once the dataset is learned, the query phase is quasi instantaneous. This is in clear departure with evolutionary methods where for each and every query the whole parameter space is searched.

To illustrate our approach, we design a novel deep network, which uses a fully connected neural network. We introduce a bi-directional deep neural network architecture composed of two networks (Fig. 2A), where the first is a Geometry-predicting-network (GPN) that predicts a geometry based on the spectra (the inverse path) and the second is a Spectrum-predicting-network (SPN) that predicts the spectra based on the nanoparticle geometry (the direct path). The geometry predicted by the GPN is fed into the SPN which, in turn, predicts back the spectrum. We thus solve the harder inverse problem first, i.e. predicting the geometry based on two spectra for both polarizations, and then, using the predicted geometry, we match the recovered spectrum with the original one (see SOM1 for further information). It is worth noting that the training of such a bidirectional network requires a dedicated learning procedure since the input to the SPN is a predicted geometry rather than the actual geometry (see SOM for more information). Furthermore, we also observe a significant gain from training one network on *all* the training sets rather than the alternative of training multiple separate networks. It is crucial to stress that the learning phase in DNN is a non recurring effort, meaning once the dataset is learned, the query phase is quasi instantaneous. This is in clear departure with evolutionary methods where for each and every query the whole parameter space is searched.

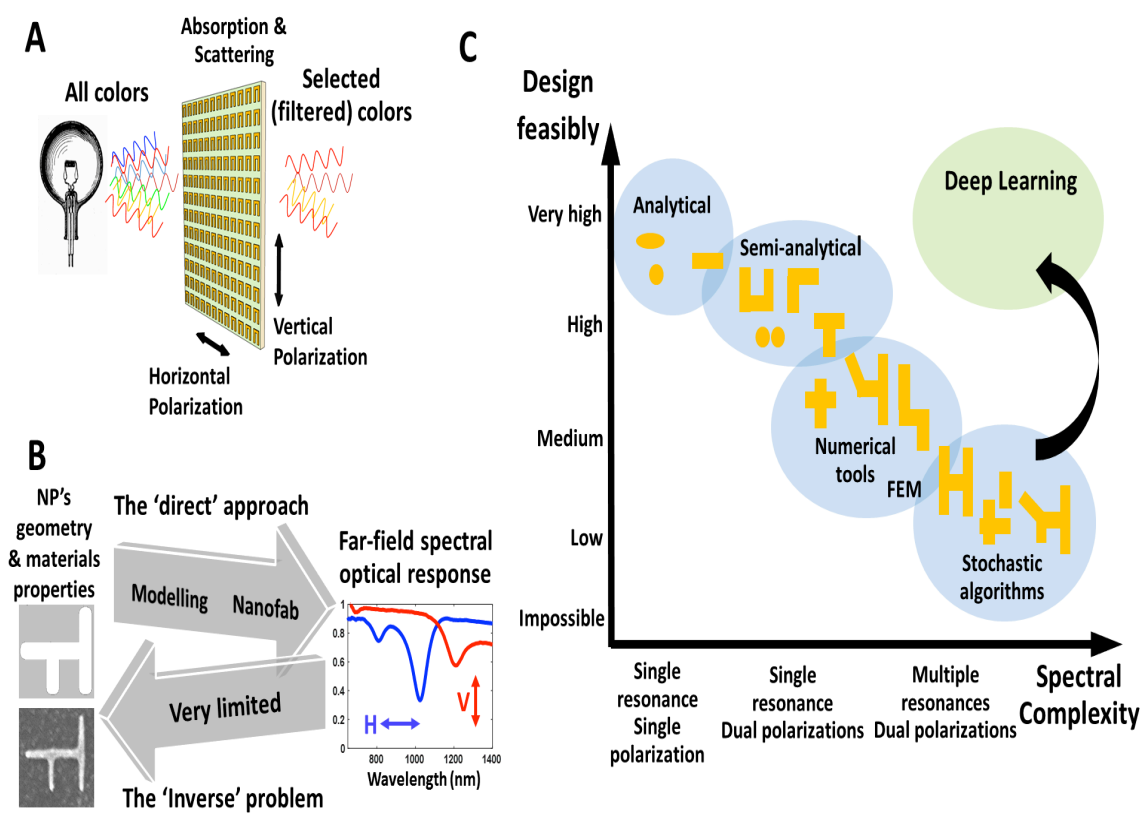

**Figure 1 - Deep learning Nano-photonics and the 'inverse' problem** (A) Interaction of light with plasmonic nanostructures. Incoming electromagnetic radiation interacts with man-made sub-wavelength structures in a resonant manner, ultimately leading to an effective optical response where the optical properties for both polarizations of the designed metamaterial are dictated by the geometry at the nanoscale rather than the chemical composition. (B) To date, the approach enabled by the computational tools available only allow for a 'direct' modeling i.e. predicting the optical response in both polarization of a nanostructure based on its geometry, constituent and surrounding media. However, the 'inverse' problem, where the tool outputs a nanostructure for a input desired optical response, is much more relevant from a designer point of view and currently unachievable in a time efficient way. The more complex the optical response desired, the more unattainable is the solution of the inverse problem (C). The deep learning approach bridges this gap and unlocks the possibility to design, at the single nanoparticle level, complex optical responses with multiple resonances and for both polarizations.

In order to train our DNN, we have created a large set of synthetic data using COMSOL Multiphysics (4). The data contains more than 15,000 experiments, where each experiment is composed of a plasmonic nanostructure with a defined geometry, its metal properties, the host's permittivity and the optical response spectrum for both horizontal and vertical polarizations of the incoming field (see Methods). While we maintained the thickness of the nanoparticle constant, it can of course influence the transmission spectra (blueshift and resonance strength). This can be added as a parameter to the learning dataset and allow refined predictions. In our proof of concept, we choose a nanostructure geometry represented by a general "H" form, where each of the outer edges can vary in length, angle or can be omitted (see Fig. 2A). Such variable geometry is complex enough to span a wide variety of optical response spectras for both polarizations. We then fed the DNN with these synthetic optical experiments and let it learn the multivariate relationship between the spectras and all of the aforementioned geometric parameters. During this training process, the prediction provided by the DNN on a set of synthetic experiments is compared to the COMSOL solutions and the network weights are optimized to minimize the discrepancy. A set of similarly created samples, unseen during training, is used to evaluate the network's performance.

We then demonstrate our DNN ability to accurately predict fabricated nanostructures' parameters beyond simulations, by fabricating a set of different geometries encompassing some which the network has never seen. We measured the transmission spectra on a home-built reflection- transmission setup, which includes a broadband source, high magnification objectives and collection optics and fiber to an Optical Spectrum Analyzer (see Methods). We fed these *measured* spectra into our trained DNN and obtained an excellent agreement between the retrieved dimensions and those actually measured by SEM (Insets Fig. 2B and 2C). These excellent predictions were obtained once the DNN was trained with a training set of 1500 simulated experiments where the network was able to learn the different geometries' response in the presence of the *measured* dispersion of the Indium Tin Oxide layer (ITO), see Methods. We emphasize that our DNN allows the retrieval of geometrical dimensions and optical properties of a subwavelength geometry from the family of the subwavelength H-geometries that reproduce its far-field spectra.

This is, to our knowledge, an unmatched capability of multivariate parameters retrieval that only such an approach has achieved. We note that this achievement is enabled by the unique bi-directional architecture and the simultaneous learning process between the GPN and SPN leading to co-adaptation between the networks. Compared to the simultaneous bi-directional training, we observe that the performance of the two separately trained GPN and SPN is significantly inferior.

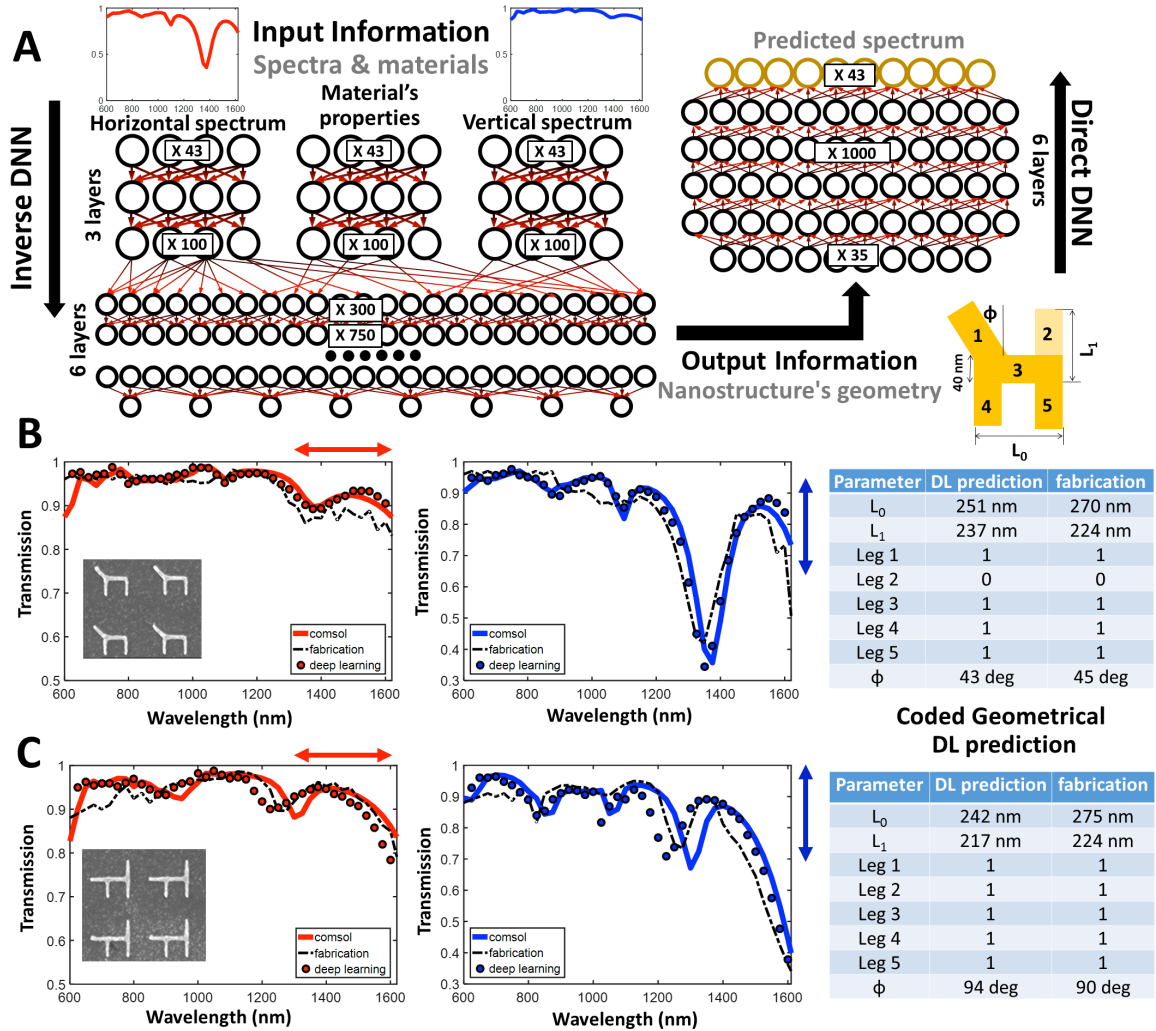

**Figure 2 – Deep learning architecture and retrieval of nanostructure's geometry.** (A) Deep networks have a cascaded structure of many layers of nonlinear processing units, where each layer uses the output from the previous layer as input. The training of our bidirectional network consists of two phases. We first train the inverse network to predict the geometry based on the transmission spectrum. In the second phase, we train the direct network on top of the first network. The inverse network receives as input parameters, two spectrums and material properties, and for each experiment it learns the corresponding geometry, material properties and resonances of the unknown geometry. The spectra are fed into the network in a raw form of 43 Y (transmission) values, where X values (wavelengths) are fixed. The first part of the inverse network is composed of three parallel pipelines that contain three layers. The second part of the inverse function consists of a sequential network that receives as input, 250 neurons from each part of the parallel network. It then joins them together and sends them to six layers. The output of this inverse network with total depth of nine layers is the predicted nanostructure geometry represented by its encoded structure, length sizes and angle. The direct network receives the predicted geometry as input and the given material properties. The output of the direct network is a transmission graph of 43 regressed values in the range [0,1], and is run twice, once for each polarization. The ReLU activation function (33) is used throughout the network. Once this DNN is trained, the nanostructure's geometry is retrieved based on the measured/desired transmission spectrum by querying the inverse network. The obtained geometry is then fed into the trained direct network which computes the predicted transmission spectrum. We also run a COMSOL simulation with the predicted geometry. The results of this process for the two different nanostructures we have fabricated are depicted in (B) and (C). The measured spectrum is depicted in a red (blue) dotted line for the horizontal (vertical) input polarization. The DL predicted geometry is represented by the different lengths in the table. The DL spectrum based on the predicted geometry is depicted as full circles. The results of the COMSOL simulations based on the DL predicted geometry are represented as full lines. For all the nanostructures the gold thickness is maintained at 40 nm.

From the novel bi-directionality nature of our network, the output of the inverse network serves as an input to the direct network and is used to predict back the two spectrums of the predicted geometry. As an example, we demonstrate the bi-directionality advantage in the case of the dispersive ITO. This advantage is apparent from the Mean Squared Error (MSE) achieved on the error function in both approaches, i.e. Bidirectional versus composite direct (SPN) & inverse (GPN) networks (more information can be found in SOM). The bidirectional network exhibits a significantly lower MSE of 0.16 compared to the MSE achieved with the composite approach (MSE=0.37). It is also worth noting an interesting property of the direct network that predicts both spectrums using the same weights. The shared weights property gives better results than two different networks with each one specializing in one polarization. This results in a significant improvement in predicting each polarization when the DNN also learns the opposite polarization. This has a physical explanation as the free electrons in the nanostructures are occupied in the same 2D boundaries, thus making the two polarizations coupled to each other.

In order to gain insight on the effect of the network's depth on the prediction performance, we conduct an extensive comparison between different network architectures. We show that different network depths have a dramatic effect on the results. We vary the number of fully connected layers at the second part of the inverse network and by comparing the results to each other, we see a significant effect on the accuracy of the prediction, as seen in Fig. 3. We find that the best inverse network architecture for our case, is three parallel group layers followed by six sequential fully connected join layers. Interestingly, we observe a significant gain in accuracy when using six join layers compared to five or seven layers in the sequential part of the network. The benefit of such a deep network is directly derived from the complexity and nonlinearity of the underlying physical process. We observe a significant gain from training one network on all of the training set over the alternative of training multiple separate networks. While this can be attributed to the so-called transfer of knowledge (34) where knowledge learned from one problem is transferred to another, we are not aware of other instances in which it is that crucial to train one single generalist network instead of applying a divide and conquer strategy with multiple specialist networks.

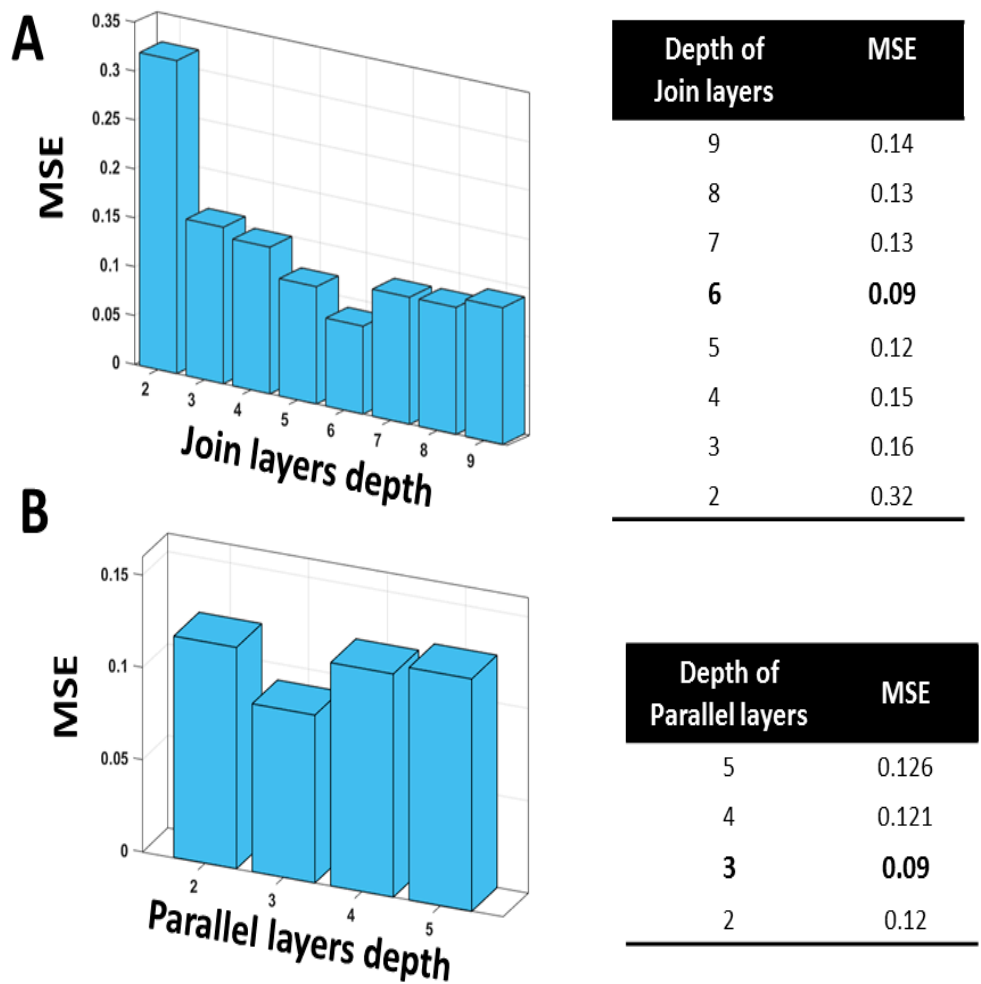

Figure 3 – **Analysis of the depth and parallelism of the network**. The performances of a trained GPN with different architectures are presented. (A) We chose the parallel layers number to be fixed and change the number of fully connected layers, namely the Join-layers depth. In this case, deeper DNN leads to better accuracy. However, there is a certain depth, in our case six layers, in which the addition of extra layers doesn't improve the results or leads to the vanishing gradient phenomena. (B) The importance of the parallel group layers part of the GPN is demonstrated. Adding parallel layers forces the network to represent each one of the three data groups separately and only then joining them together to one big inner representation. Using parallel group layers improves the results but the number of them should be chosen carefully.

Next, we have examined the strength of the inverse predictive approach for sensing applications where plasmonic nanostructures are used for enhance the light-matter interaction with various chemicals and bio-molecules. Organic compounds typically exhibit pronounced resonances across the spectrum from ultraviolet to mid-infrared. We show that our trained DNN allows us to find the nanostructure configuration to best interact with a given molecule with target multiple resonances in the two polarizations. More specifically, we wish to design a nanostructure targeted to enhance the interaction with Dichloromethane, an important chemical used in industrial processes. This organic compound exhibits one resonance around 1150 nm and another around 1400-1500 nm. Our design goal is to achieve a nanostructure that will resonate in a water solution (at both wavelengths for one polarization and will have completely different resonances at the orthogonal polarization, at around 820 nm (match a Ti:Sapphire femtosecond laser excitation for a pump-probe experiment), 1064nm and 1550nm. In the existing design process, this task would require it to iterate through different designs using the standard FEM or FDTD simulation tools, a process that can be extremely time consuming. The DNN's inverse solution yields, in few seconds, the parameters shown in Fig. 4 (C). We also applied this design approach to the asymmetrical Phthalocyanine dimer 1a, a synthetic molecule which has more complex polarization characteristics (Fig 4(D-E)) with potential applications due to its charge transfer properties (35). The DNN inverse design for this targeted molecule and polarizations results in the configuration shown in Fig. 4(E). This demonstrates the capability of our DNN to address various targeted resonances in different polarizations and emphasizes that this approach can be extended to other molecules for sensing in biology, chemistry and material science.

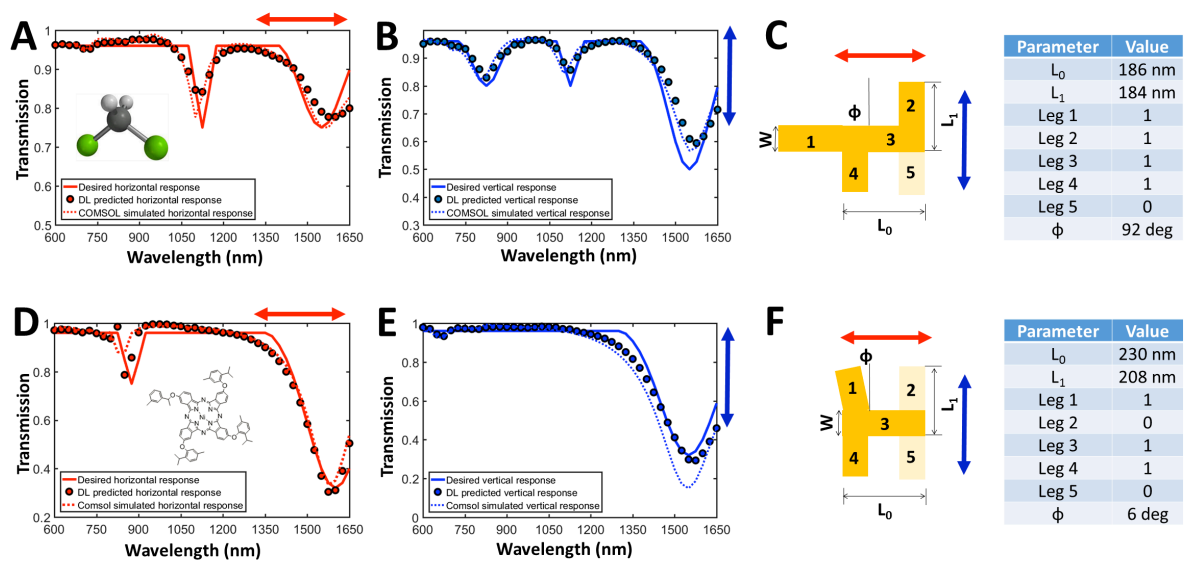

Figure 4 – **Prediction of the nanostructure's geometry for chemical sensing.** (A-C) DNN based design of a gold plasmonic structure targeted to the organic molecule Dichloromethane with different spectral polarization response (A-B) on one polarization axis. It has two resonances on 1150nm and a broad one between 1400-1600nm, whereas on the orthogonal polarization, it has three resonances, at around 820 nm (match a Ti:Sapphire femtosecond laser excitation for a pump-probe experiment), 1064nm and 1550nm (C) Configuration and dimensions of the plasmonic structure found by the DNN. (D-E) The targeted molecule is asymmetrical Phthalocyanine dimer 1a, a synthetic molecule which has more complex polarization characteristics, with potential applications due to its charge transfer properties. (F) Configuration and dimensions of the plasmonic structure found by the DNN. In both geometries, after

prediction of the geometry, COMSOL simulations were performed, showing an excellent agreement with the desired spectra. This design approach can be extended to other molecules for biology, chemistry or material sciences.

To conclude, in this work, we design, train and test a novel deep-learning architecture, showing a capability for predicting the geometry of nanostructures based solely on the far-field response of the nanostructures. The proposed scheme allows a very accurate prediction of the geometry of a complex nanostructure and could be extended to other physical and optical parameters of the host materials and compounds. The approach also effectively addresses the currently inaccessible inverse problem of designing a geometry for a desired optical response spectrum and also significantly speeds up the direct spectrum prediction of such sub-wavelength structures. This approach breaks the ground for the on-demand design of optical response for many applications such as sensing, imaging and also for Plasmon's mediated cancer thermotherapy. We believe that artificial intelligence in general and deep learning approaches in particular have far-reaching implications in the field of nanotechnology and nanophotonics.

# Methods

Sample Preparation

ITO covered glass (Sigma Aldrich) were covered with PMMA-A4 polymer and spin-coated for one minute at 7,000 RPM. The electron beam (Raith150) used was 10kV beam, aperture, 6mm WD and a dose was deposited in single-pixel lines. Samples were then developed in MIBK/IPA (1:3) for 1 minute and rinsed in isopropanol for 20 seconds. of gold were then evaporated on the sample with E-Beam Evaporator (VST evaporator). Lift-off was done with acetone and followed with a final wash in isopropanol.

Sample Characterization

Sample sizes were verified using an electron microscope and were optically characterized using an OSL2 Broadband Halogen Fiber Optic Illuminator (Thorlabs) light-source and LPNIR050 (Thorlabs) broad band polarizer. Transmitted light was filtered in an imaging plane by an iris such that only light which passed through the sample was collected and then analyzed by a AQ6370D (Yokogawa) spectrometer.

COMSOL Simulation

We performed finite element method (FEM) simulations using the 'Electromagnetic Waves, Frequency Domain' module of the COMSOL 4.3b commercial software was utilized. For consistency, the edges were made using fillets with the constant radius of 15nm. We have considered geometries based on a five edges shape of 'H' while varying an angle of one of the edges, the existing edges and the edges lengths.

The nanostructure is simulated in a homogeneous dielectric medium with a chosen real effective-permittivity. For preventing reflections from the far planes, PMLs with a depth of the maximum wavelength were placed on both far ends of the homogeneous medium in the propagation direction of the radiating field.

For the dataset predicting the fabrications, the nanostructure was made of Gold with a wavelength dependent homogeneous medium permittivity () such that [S1], where stands for the air permittivity and equals 1 and is the ITO permittivity and is wavelength dependent such that its imaginary part can be neglected in the measured spectrum range. A justification for ignoring the glass permittivity can be found in [S2] and [S3]. In [S2] it was shown that changes in the thickness of a Titanium adhesion layer higher than 40% of the nanostructures height, doesn't affect the plasmon resonance. In [S3] it was shown that for an Au nanoparticle with diameter of 10nm and a graphene layer, the LSPR shifting saturates when the distance is more than 20 nm.

A prediction for a similar behavior of the ITO layer is assumed. In our case, the ITO thickness is about 100nm which is about 250% of the nanostructure thickness of about 40nm.

# References

1. N. Yu, F. Capasso, "Flat optics with designer metasurfaces" Nature Materials 13, 139–150 (2014).

2. A. V. Kildishev, A. Boltasseva, V. M. Shalaev, "Planar photonics with metasurfaces" Science 339, 1232009–1232009 (2013).
3. X. Ni, Z. J. Wong, M. Mrejen, Y. Wang, X. Zhang, "An ultrathin invisibility skin cloak for visible light" Science 349, 1310–1314 (2015).
4. COMSOL Multiphysics v. 5.2. www.comsol.com. COMSOL AB, Stockholm, Sweden.
5. A. F. Oskooi, D. Roundy, M. Ibanescu, P. Bermel, J. D. Joannopoulos, S. G. Johnson, "MEEP: A flexible free-software package for electromagnetic simulations by the FDTD method" Computer Physics Communications 181, 687–702 (2010)
6. Lumerical Solutions, Inc. http://www.lumerical.com/tcad-products/fdtd/
7. D. Colton, R. Kress, "Inverse acoustic and electromagnetic scattering theory." 93, Springer New York (2013).
8. T. W. Odom, E.-A. You, C. M. Sweeney, "Multiscale plasmonic nanoparticles and the inverse problem" J. Phys. Chem. Lett. 3, 2611–2616 (2012).
9. Nobel Prize in Chemistry announcement by the Royal Swedish Academy of Sciences for 2014 Available at: https://www.nobelprize.org/nobel_prizes/chemistry/laureates/2014/advanced-chemistryprize2014.pdf
10. E. Betzig, et al. "Imaging intracellular fluorescent proteins at nanometer resolution" Science 313, 1642–1645 (2006).
11. M. J. Rust, M. Bates, X. Zhuang, "Sub-diffraction-limit imaging by stochastic optical reconstruction microscopy (STORM)" Nat Meth 3, 793–796 (2006).
12. S. T. Hess, T. P. K. Girirajan, M. D. Mason, "Ultra-high resolution imaging by fluorescence photoactivation localization microscopy" Biophysical Journal 91, 4258–4272 (2006).
13. D. Macías, P.-M. Adam, V. Ruiz-Cortés, R. Rodríguez-Oliveros, J. Sánchez-Gil, "Heuristic optimization for the design of plasmonic nanowires with specific resonant and scattering properties," Opt. Express 20, 13146–13163 (2012).
14. G. M. Sacha, P. Varona, "Artificial intelligence in nanotechnology", Nanotechnology 24 452002 (2013).
15. P. Ginzburg, N. Berkovitch, A. Nevet, I. Shor, M. Orenstein, "Resonances on-demand for plasmonic nano-particles", Nano Lett. 11, 2329–2333 (2011)
16. Forestiere et al, Particle-swarm optimization of broadband nanoplasmonic arrays. Opt. Lett. 35, 133 (2010);
17. Genetically engineered plasmonic nanoarrays, Nano Lett. 12, 2037 (2012);
18. Feichtner et al, Evolutionary optimization of optical antennas. Phys. Rev. Lett. 109, 127701 (2012);
19. Forestiere et al, Inverse design of metal nanoparticles morphology, ACS Photonics 1, 68 (2015)."
20. A. Krizhevsky, I. Sutskever, G. E. Hinton, "ImageNet classification with deep convolutional neural networks". In NIPS, 1106–1114 (2012).
21. G. Hinton, L. Deng, D. Yu, G. Dahl, A. Mohamed, N. Jaitly, A. Senior, V. Vanhoucke, P. Nguyen, T. Sainath, B. Kingsbury, "Deep neural networks for acoustic modeling in speech recognition" IEEE Signal Processing Magazine, 2982–97 (2012).
22. R. Socher, A. Perelygin, J. Wu, J. Chuang, C. D. Manning, A. Y. Ng, C. Potts, "Recursive deep models for semantic compositionality over a sentiment treebank" In Proceedings of the 2013 Conference on Empirical Methods in Natural Language Processing, pages 1631–1642, Stroudsburg, PA (2013).
23. Y. Taigman, M. Yang, M. Ranzato, L. Wolf, "Deepface: closing the gap to human-level performance in face verification" In Proc. Conference on Computer Vision and Pattern Recognition 1701–1708 (2014).
24. P. Baldi, P. Sadowski, and D. Whiteson , "Searching for exotic particles in high-energy physics with Deep Learning", Nat. Comm. 5, 4308 (2014).
25. P. B. Wigley, P. J. Everitt1, A. van den Hengel, J. W. Bastian, M. A. Sooriyabandara, G. D. McDonald, K. S. Hardman, C. D. Quinlivan, P. Manju, C. C. N. Kuhn, I. R. Petersen,A. N. Luiten, J. J. Hope, N. P. Robins & M. R. Hush, ""Fast machine-learning online optimization of ultra-cold-atom experiments", Scientific Reports 6, 25890 (2016).
26. W. J. Brouwer, J. D. Kubicki, J. O. Sofo, C. L. Giles, "An investigation of machine learning methods applied to structure prediction in condensed matter", arXiv:1405.3564 (2014).
27. K. Hansen, et al. "Assessment and validation of machine learning methods for predicting molecular atomization energies" J. Chem. Theory Comput. 9, 3404–3419 (2013).
28. L. Waller, L. Tian, "Machine learning for 3D microscopy", Nature 523, 416-417 (2015).
29. C. L. Chen, A. Mahjoubfar, L. C. Tai, I. K. Blaby, A. Huang, K. R. Niazi, B. Jalali, "Deep Learning in label-free cell classification", Scientific Reports 6, 21471 (2016).
30. W. Cai, V. Shalaev, "Optical metamaterials" Springer New York. doi:10.1007/978-1-4419-1151-3 (2010).
31. P. Latimer, "Light scattering by ellipsoids" J. of Colloid and Interface Science 53, 102–109 (1975).

32. R. R. Oliveros, R. P. Domínguez, J. A. S Gil, D. Macías, "Plasmon spectroscopy Theoretical and numerical calculations, and optimization techniques", Nanospectroscopy 1:67–96 (2015).
33. RelU - X. Glorot, A. Bordes, Y. Bengio, "Deep sparse rectifier neural networks", Proc. Conf. Artificial Intelligence and Statistics (2011).
34. J. Yosinski, J. Clune, Y. Bengio, H. Lipson, Z. Ghahramani, M. Welling, C. Cortes, N. Lawrence, K. Weinberger, "How transferable are features in deep neural networks?" in Advances in Neural Information Processing Systems 2, pp. 3320-3328, Curran Associates, Inc. (2014).
35. G. Huang, et al. ''A series of asymmetrical Phthalocyanines: synthesis and near Infrared properties", Molecules, 18, 4628–4639 (2013).
Author Contributions:

H.S. conceived the project. M.M. and A. N. conducted the COMSOL simulations. I. M. and L. W. designed and implemented the Deep Learning Network. U.A. fabricated and characterized the nanostructures. . All authors discussed the results and wrote the manuscript.

**Competing financial interests**

The authors declare no competing financial interests.